\documentclass[twocolumn,showpacs,preprintnumbers,prl]{revtex4} 
\usepackage{graphics}
\begin{document}
\title{Out-of-equilibrium states as statistical equilibria of an effective dynamics}
\author{
Julien Barr{\'e}$^{1,2}$\thanks{Julien.Barre@ens-lyon.fr}, Freddy Bouchet$^{2}$\thanks{bouchet@poincare.dma.unifi.it}, Thierry
Dauxois$^{1}$\thanks{Thierry.Dauxois@ens-lyon.fr},  Stefano
Ruffo$^{1,2}$\thanks{ruffo@ingfi1.ing.unifi.it}}
\affiliation{1. Laboratoire de Physique, UMR-CNRS 5672, ENS Lyon, 46
All\'{e}e d'Italie, 69364 Lyon c\'{e}dex 07, France\\
2.Dipartimento di Energetica, Universit{\`a} di Firenze,INFM and INFN,
via S. Marta, 3, 50139 Firenze, Italy}

\date{\today}

\begin{abstract}
  We study the formation of coherent structures in a system with
  long-range interactions where particles moving on a circle interact
  through a repulsive cosine potential.  Non equilibrium structures
  are shown to correspond to statistical equilibria of an effective
  dynamics, which is derived using averaging techniques. This simple
  behavior might be a prototype of others observed in more complicated
  systems with long-range interactions, like two-dimensional
  incompressible fluids  and wave-particle interaction in a plasma.
\end{abstract}

\maketitle

The behavior of many complex nonlinear dynamical systems results in
the formation of spatially ordered structures.  Examples include
two-dimensional incompressible weakly dissipative
fluids~\cite{hasegawa}, two and three-dimensional
magnetohydrodynamics~\cite{Carnevale}, planet
atmospheres~\cite{freddyJFM}, electrostatic potential in inhomogeneous
magnetized plasmas, galaxy cluster formation in self-gravitating
astrophysical systems~\cite{Padmanabhan}, vortices in rotating
Bose-Einstein condensates~\cite{BEC}.  Similar effects were also
numerically revealed~\cite{dauxoispeyrard} in discrete lattices, in
which the modulational instability of a linear wave was shown to be
the first step towards energy localization, followed by the nonlinear
interaction among breather-like excitations.  Related phenomena have
been experimentally reported, such as the appearance of oscillons in
vibrated granular media~\cite{swinney}.

Besides all analogies, the reasons for this {\it self-organization}
can be indeed quite different.  In systems for which a structure
originate from any randomly chosen initial condition, a natural
explanation is of a statistical nature: see for instance
Ref.~\cite{lyndenbell} for self-gravitating systems or
Ref.~\cite{robert} for two dimensional conservative fluids. In other
systems, the structures arise and persist as a direct consequence of
the nonlinear dynamics as for instance in Ref.~\cite{dauxoispeyrard}.
We present in this letter a simple and fully tractable model for which
both reasons for self-organization, statistical mechanics and
nonlinear dynamics, have to be invoked. Indeed, we will show that
Gibbsian statistical mechanics can be safely applied {\em only} when
the correct dynamical variables are singled out~\cite{Elskens}.

The model, with long range forces, exhibits long lived out of
equilibrium structures and related unusual relaxation
phenomena~\cite{Latora}, like those encountered in self-gravitating
systems, plasmas or two dimensional fluids. Despite the importance of
such systems, a clear understanding of their peculiarities is still an
open problem.

\begin{figure}
\resizebox{0.48\textwidth}{!}{\includegraphics{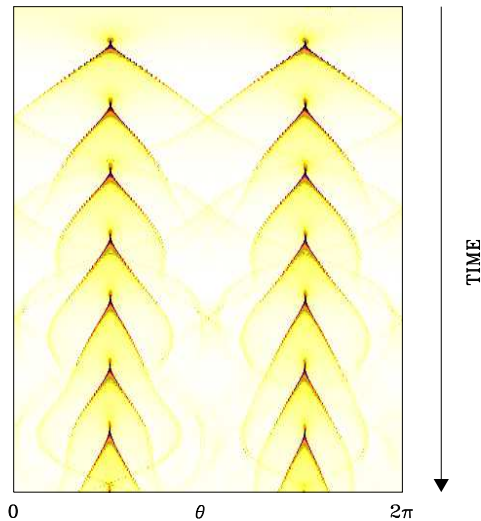}}
\vskip -1truecm
\caption{Bicluster formation : short-time evolution of the particle density
in grey scale: the darker the grey, the higher the density. Starting from
an initial condition with all the particles evenly distributed on the circle,
one observes a very rapid concentration of particles, followed by the
quasi periodic appearence of ``chevrons'', that shrink as time increases. The
structure later stabilizes and form two well-defined clusters.}
\label{chevrons1}
\end{figure}

We study the repulsive Hamiltonian Mean Field (HMF)
model~\cite{Antoni}, that describes the motion of $N$ particles on a
circle. Its Lagrangian is
\begin{eqnarray}\label{completelagrangian}
L\left( {\theta_i},{\dot{\theta_i}}\right) & = & \sum_{i=1}^{N}
\frac{\dot{\theta_i}^2}{2}
-\frac{1}{2N}\sum_{i,j=1}^N\cos(\theta_i-\theta_j)\quad.
\end{eqnarray}
This model is an archetype of mean field models for which interactions are
infinite-range with size dependent coupling~\cite{Kac}.  For a special but wide
class of initial conditions, this model~\cite{drh,lfr} has a very interesting
dynamics since, contrary to statistical mechanics expectations, a
localized structure (bicluster) appears (see Fig.~\ref{chevrons1}).
Its initial formation results from a fast oscillation of the medium,
that nonlinearly drives, on a longer time scale, an average motion of
the particles in an effective double-well potential. The mechanism is 
qualitatively reminiscent of the parametrically forced pendulum, first 
analyzed by Kapitza~\cite{kapitza}: the role of the harmonic forcing 
of the pivot  is here played, by the \emph{self induced} 
fast oscillation of the original dynamics. As for the Kapitza pendulum,  
we average over the fast oscillation
and obtain an effective Hamiltonian  for the slow motion. We
will show that the final bicluster structure is a statistical
equilibrium of such an effective Hamiltonian.

Let's consider the evolution of particles for small initial energy and
velocity dispersion~\cite{drh}.  The analysis relies on the existence
of two time-scales in the system: the first one, defined by the
linearized dynamics, is intrinsic and corresponds to the inverse of
the ``plasma frequency" $\omega=\sqrt2/2$.  The second one depends on the
energy per particle $e=H/N$ via the relation $\tau=\varepsilon t$, with $\varepsilon=
\sqrt{2e}$.  As a consequence, we use the following ansatz
$\theta_i=\theta_i^0(\tau)+\varepsilon f_i(t,\tau)$.  A multi scale analysis shows that the
correct choice to describe the small and rapid oscillations is $f_i=
\left[\sqrt{2} A_+(\tau)\sin(\phi_+(t))+a_+(\tau)\right]\ \cos(\theta_i^0+\psi) +
\left[\sqrt{2}A_-(\tau)\sin(\phi_-(t))+a_-(\tau)\right]\ \sin(\theta_i^0+\psi)$ where
$\phi_\pm(t)$ are fast variables, and $d\phi_\pm/dt$, $A_\pm(\tau)$, and $a_\pm(\tau)$ are
slow variables.  The phase $\psi$ is defined via the complex number
  $M_2=N^{-1}\sum_{j=1}^{N} e^{2i\theta_j^0}\equiv|M_2|e^{-2i\psi}$.  We introduce
this expression in Lagrangian~(\ref{completelagrangian}) and keep
terms up to order $\varepsilon^2$, averaging over the fast time $t$, as
described e.g. in Ref.~\cite{witham}.  Dropping the $\varepsilon^2$ overall
factor, the averaged Lagrangian reads
\begin{equation}
L_{slow} = \frac{1}{2}\sum_{i=1}^N \left(\frac{d\theta_i^0}{d\tau}\right)^2
+L_+ + L_-~,
\label{Lagrangianslow}
\end{equation}
where
\begin{eqnarray}
L_{\pm}&=&\frac{N}{2}\Bigl[A_\pm^2{\dot \phi_\pm}^2 \omega_\pm^2 
-(M_{1\pm}^0)^2 \mp 2 M_{1\pm}^0 a_\pm\omega_\pm^2 \nonumber \\
&&-\left({A_\pm^2}+{a_\pm^2}\right)\omega_\pm^4\Bigr]~.
\label{elleplusmoins}
\end{eqnarray}

$M^0_1 \equiv \varepsilon N^{-1}\sum_{j=1}^{N} e^{i\theta_j^0}=M_{1-}^0+iM_{1+}^0$ 
is the first moment associated to the angles and $\omega_{\pm}\equiv\sqrt{(1\pm |M_2|)/2}$ 
will turn out to be the fast frequencies (for details see Ref.~\cite{bbdr}).
$\phi_\pm$ are now cyclic variables and correspond to two
quasi-conserved quantities, $P_\pm={NA_\pm^2\omega_{\pm}^2\dot \phi_\pm}$, 
which are adiabatic invariants of the dynamics. Going from the 
Lagrangian to the Hamiltonian formalism, one must
first remark that, having eliminated the dependence on the time
derivatives of $A_\pm$, $a_\pm$, there is no Legendre transform over these
variables. In the absence of conjugate
variables to the amplitudes $A_\pm$, the corresponding Hamilton's
equation are simply given, from the least action principle, by
$\partial_{A_\pm} H_{slow}=0$.  This leads to the expression for the amplitudes
$NA_\pm^2 = P_\pm\omega_{\pm}^{-3}$.  Finally, from the equations
$\partial_{a_\pm} H_{slow} =0$ and $\dot \phi_\pm =\partial_{P_\pm} H_{slow}$,
 we determine the shifts $a_{\pm}$ and the phases $\phi_\pm$ .

Rewriting the action, taking into account the relations for $A_\pm$ and $a_{\pm}$, we 
end up with a very simple \emph{effective} Hamiltonian that retains 
only the slow motion
\begin{equation}
H_{eff} = \sum_i \frac{{p_i^0}^2}{2}
+P_+\sqrt{\frac{1+|M_2|}{2}} + P_-\sqrt{\frac{1-|M_2|}{2}}~,
\label{heff}
\end{equation}
and which describes the evolution of the full system. The constant $P_+$ and
$P_-$ are determined by the initial condition. 

\begin{figure}
\resizebox{0.42\textwidth}{!}{\includegraphics{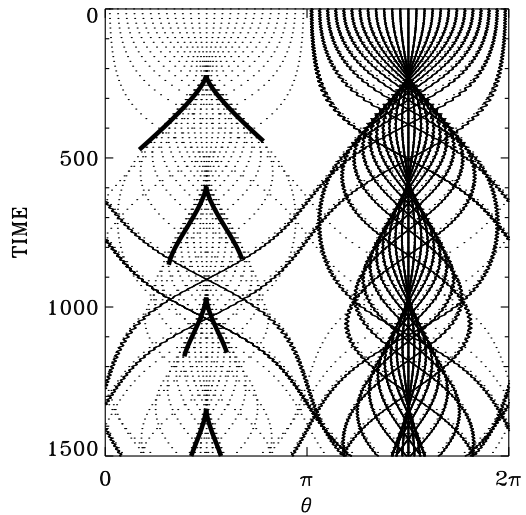}}
\vskip -.5truecm
\caption{Spatio-temporal evolution of 50 particles with initial energy 
  per particle $e=2.5\ 10^{-5}$.  The dotted lines present the results
  given by the effective Hamiltonian (\ref{heffP}) whereas the solid
  lines show the evolution of particles initially between $\pi$ and $2\pi$
  according to the original Lagrangian (\ref{completelagrangian}):
  except for the fast small scales oscillations, they are almost
  indistinguishable. On the left (between 0 and
  $\pi$), the caustics (thick lines) given by formula (\ref{chevrons}) 
reproduce the ``chevrons''.}
\label{compatrajprl}
\end{figure}

For the sake of simplicity, we restrict here to the case in which
only one mode, $P_-$, is initially excited ($P_+=0$). 
Dropping the subscript for $P$, we then consider the
Hamiltonian 
\begin{eqnarray}
H_{eff} = \sum_i \frac{{p_i^0}^2}{2}+ P\sqrt{\frac{1-|M_2|}{2}}\quad.
\label{heffP}
\end{eqnarray}
The particle dynamics determined by this Hamiltonian perfectly
compares with the one given by Lagrangian (\ref{completelagrangian}), as
shown in Fig.~\ref{compatrajprl}. 

In order to explain the formation of the bicluster and the 
chevrons observed in Fig.~\ref{chevrons1}, let us first remark that the
equations of motion of the effective Hamiltonian $H_{eff}$
\begin{equation}
\label{equationsofmotion}
\ddot{\theta}_i^0=-\frac{P}{N\sqrt{2(1-|M_2|)}}\sin{2(\theta_i^0+\psi)}
\end{equation}
are pendulum-like.  The evolution of the system thus consists of $N$
pendulum trajectories in a double-well potential, which is {\em
  self-consistently} modified since $M_2$ depends on the ${\theta_i^0}$.
If the initial velocity dispersion is small, singularities appear in
the density profile which becomes infinite along the envelope (or
caustics) of the trajectories; the chevrons of Fig.~\ref{chevrons1}
are a manifestation of these singularities~\cite{ArnoldZeldovich}.
This picture allows us also to make quantitative predictions.  To
simplify the calculations, we assume that position and depth of the
double well potential are fixed; this amounts to take $\psi=const$ in
Eq.~(\ref{equationsofmotion}) (as suggested by Fig.~\ref{chevrons1})
and $|M_2|=const$ which is the simplest hypothesis we can
make in order to characterize the chevrons.  We can compute the time
$t_s$ of the first divergence of the density: it is the time when the
particles in the bottom of the two wells of the potential, that
perform a quasi-harmonic motion, cross. We get
$t_s=\pi\sqrt{N\omega_-/(8Pe)}$, where $P$ and $e$ depend on the initial
condition, and $\omega_-$ depends on $|M_2|$.  Approximating the
trajectories by straight lines near the bottom of the wells, we obtain
the shape of the ``chevrons'', resorting to standard
methods~\cite{Zauderer} of curve envelope calculation (see details
in~\cite{bbdr}).  The lowest order approximation of the $n-$th shock
is
\begin{eqnarray}
\label{chevrons}
\theta & \propto & \frac{\left(t-(2n-1)t_s \right)^{3/2}}{\sqrt{2n-1}}~,
\end{eqnarray}
where the $1/\sqrt n$ factor explains the shrinking of the chevrons,
whereas the $3/2$ scaling law is generic according to catastrophe's
theory~\cite{ArnoldZeldovich}. Taking for $|M_2|$ its mean value
($0.51$ for the present initial condition), Fig.~\ref{compatrajprl}
emphasizes that the agreement with the numerics is quantitatively
excellent.


Since the short time dynamics and the ``chevrons'' formation is now
clarified, let us next consider the stabilization of this out of
equilibrium state: this is where statistical mechanics comes into play.

\begin{figure}
\resizebox{0.42\textwidth}{!}{\includegraphics{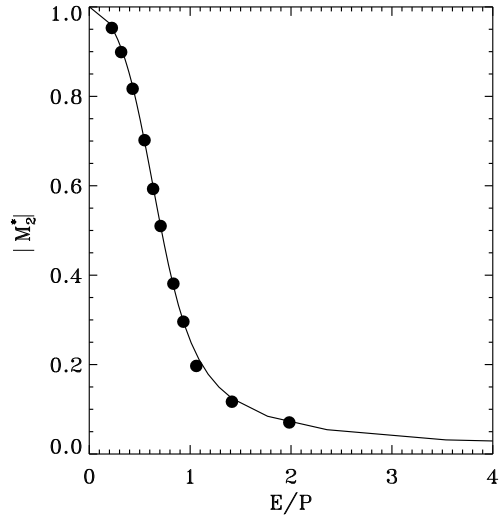}}
\vskip -.5truecm
\caption{The solid curve shows the theoretical prediction of $|M_2^{\ast}|$ as a 
  function of the ratio between the energy $E$ and the adiabatic
  invariant $P$, whereas the filled circles correspond to numerical
  results for the full Lagrangian~(\ref{completelagrangian}).}
\label{solutionthermoprl}
\end{figure}

Due to the special form of the effective potential,
which depends only on the global variable $|M_2|$, the statistical
mechanics of the effective Hamiltonian (\ref{heffP}) is exactly tractable.
Indeed, the density of states $\Omega(E)$ can be expressed in
terms of the density of states corresponding to the 
kinetic part of the Hamiltonian $\Omega_{kin}$ and of
the angular configurational part $\Omega_{conf}$  as follows
\begin{eqnarray}
\Omega(E) 
& \propto & \int d|M_2| \: \Omega_{kin}\left(E-V(|M_2|)\right) \:
          \Omega_{conf}(|M_2|)\qquad
\label{OmegaE}
\end{eqnarray}
where $V(|M_2|)=P\sqrt{(1-|M_2|)/2}$.  Using an inverse Laplace transform and the
Hubbard-Stratonovitch trick, one also obtains $\Omega_{conf}$  after some
 calculations.
$\Omega(E)$ in Eq.~(\ref{OmegaE}) can thus be evaluated~\cite{bbdr} by the
saddle point method, which leads to the
equilibrium value $|M_2^{\ast}|$ as a function of $E/P$.  This explains
the numerical result $|M_2|\simeq 0.5$ found by the authors of
Ref.~\cite{drh} (their initial conditions correspond to
$E/P=\sqrt{2}/2$ and $|M_2^{\ast}|=0.51$).  In addition, this shows that
other initial conditions lead to other values of $|M_2^{\ast}|$, which
are in excellent agreement with numerical simulations (see
Fig.~\ref{solutionthermoprl}).  The long lifetime of these
out-of-equili\-brium states is therefore fully understood, since they
appear as equilibrium states of an effective Hamiltonian that well
represents the long-time motion.

\begin{figure}
\resizebox{0.42\textwidth}{!}{\includegraphics{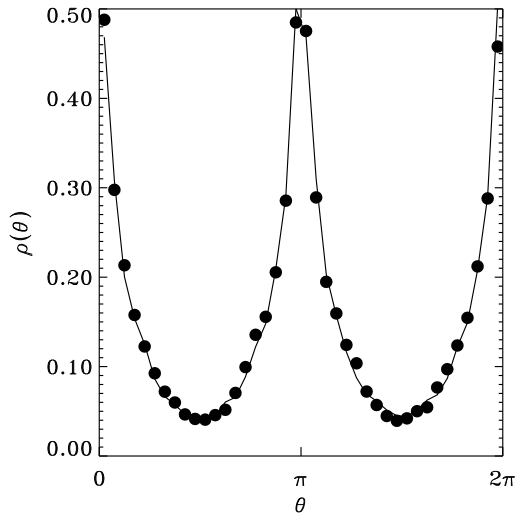}}
\vskip -.5truecm
\caption{Comparison of the non Gibbsian  density distributions 
$\rho(\theta)$ of particles obtained with the original Lagrangian~(\ref{completelagrangian}) 
(circles) and with the effective one (\ref{heffP}) (solid line).
Both results have been obtained for $N=10^3$ particles and are averaged over 
times corresponding to $\tau=10^3 \to 10^4$.
The energy per particle is  $e=2.5\ 10^{-5}$.} 
\label{compahisto}
\end{figure}

Once $|M_2^{\ast}|$ is known, it is easy to show that the equilibrium
velocity distribution is Maxwellian with temperature $T={2}\langle E_c
\rangle/{N}$, and that the distribution of angles has a Gibbsian shape
$\rho(\theta)\propto e^{-V(\theta)/T}$, with the potential inferred from the equations of
motion.  However, whereas $|M_2|$ quickly reaches its equilibrium
value, the distribution $\rho(\theta)$ relaxes very
slowly, in the numerical experiments.  Moreover, the relaxation time
increases with $N$.  Nevertheless, the density profile $\rho(\theta)$ obtained
in long-time simulations with the full
system~(\ref{completelagrangian}) fully agrees with the one obtained
with the effective dynamics~(\ref{heffP}), as shown in
Fig.~\ref{compahisto}.  This makes this model a good candidate to
study the unusual relaxation properties and the non equilibrium states
observed in other systems with long range interactions~\cite{Latora}.
  
We thus conclude that the dynamics of the effective
Hamiltonian~(\ref{heffP}) parameterizes very well the one of the full
Lagrangian~(\ref{completelagrangian}), for short as well as for long
time. The variational multi-scale analysis we used allows us to
exhibit naturally the adiabatic invariants and to preserve the
Hamiltonian structure of the problem (to leading order), making it
well suited for a statistical treatment.  This leads us to predict
statistical properties of the full system, as the asymptotic value of
$|M_2|$. Moreover, the effective Hamiltonian gives the opportunity to
study numerically the relaxation towards equilibrium of the bicluster,
whereas it was not possible in the real dynamics, because of
computational limitations. Indeed, let us observe that the ratio of
the typical time scale of the two dynamics is of order 100 or even
larger at smaller energies.

We have thus described a simple mechanism to explain the existence of
a stable out-of-equilibrium structure in a Hamiltonian mean-field
model.  This model deserves special attention for different reasons.
It is probably the simplest $N$-particles solvable model in one
dimension which exhibits such a stabilization effect, corresponding to
the coupling of very fast oscillations self interacting with a slower
motion. This model is moreover a simple analogue of other examples of
nonlinear interactions of rapid oscillations with a slower global
motion like the piston problem~\cite{piston}: averaging technics
could be applied to the fast motion of gas particles in a piston which
itself has a slow motion~\cite{sinairusse}. Examples can also be
found in applied physics as for instance wave-particles interaction in
plasma physics~\cite{plasma}, or the interaction of fast inertia
gravity waves with the vortical motion for the
rotating Shallow Water model~\cite{embidmajda}.

  We acknowledge useful discussions with D. Escande, Y. Elskens, M.C.
  Firpo, J.L. Lebowitz, F. Leyvraz. This work has been 
  supported by the French Minist{\`e}re de la Recherche by a Lavoisier
  fellowship and with grant ACI jeune chercheur-2001 N$^\circ$ 21-31, the
  R{\'e}gion Rh{\^o}ne-Alpes for the fellowship N$^\circ$ 01-009261-01. It is
  also supported by the EU contract No.  HPRN-CT-1999-00163 (LOCNET
  network) and the contract COFIN00 on {\it Chaos and localization in
    classical and quantum mechanics}.

\end{document}